\documentclass[5p,twocolumn]{elsarticle}  

\usepackage{graphicx}  
\usepackage{graphics}
\usepackage{dcolumn}   
\usepackage{bm}        
\usepackage{amssymb}   
\usepackage{amsmath}   
\usepackage{caption}
\usepackage{epsfig}
\usepackage{color}
\usepackage{multirow}

\begin{document}
\begin{frontmatter}

\title{
Pion electroproduction measurements in the nucleon resonance
region}

\address[temple]{Temple University, Philadelphia, PA 19122, USA}
\address[jlab]{Jefferson Lab, Newport News, VA 23606, USA}
\address[NMSU]{New Mexico State University, Las Cruces, NM 88003, USA}
\address[florida]{Florida International University, University Park, Florida 33199, USA} 
\address[5]{Catholic University of America , Washington, DC 20064.} 
\address[6]{Hampton University , Hampton, VA 23669.} 
\address[7]{Mississippi State University, Miss. State, MS 39762.} 
\address[8]{The College of William and Mary, Williamsburg, VA 23185.} 
\address[9]{Old Dominion University, Norfolk, VA 23529.} 
\address[10]{University of Regina, Regina, SK S4S 0A2, Canada.} 
\address[11]{Argonne National Laboratory, Lemont, IL 60439.} 
\address[12]{Artem Alikhanian National Laboratory, Yerevan, Armenia.} 
\address[13]{University of Tennessee, Knoxville, TN 37996.} 
\address[14]{Veer Kunwar Singh University, Arrah, Bihar 802301, India.} 
\address[15]{University of Pavia, 27100 Pavia PV, Italy.} 
\address[16]{Duke University, Durham, NC 27708.} 
\address[17]{University of Virginia, Charlottesville, VA, 22904.} 
\address[18]{INFN, 27100 Pavia (PV), Italy.} 
\address[19]{Virginia Polytechnic Institute \& State University, Blacksburg, Virginia 24061, USA.}

\author[temple]{R.~Li}

\author[temple]{N.~Sparveris\corref{cor1}}
\ead{sparveri@temple.edu} \cortext[cor1]{Corresponding author.}

\author[temple]{H.~Atac}

\author[jlab]{M.~K.~Jones}

\author[NMSU]{M.~Paolone}

\author[17]{Z. Akbar}
\author[NMSU]{M.~Ali}
\author[8]{C. Ayerbe Gayoso}
\author[5]{V. Berdnikov}
\author[6,19]{D.Biswas}
\author[19]{M. Boer}
\author[jlab]{A. Camsonne}
\author[jlab]{J. -P. Chen}
\author[jlab]{M. Diefenthaler}
\author[temple]{B. Duran       }
\author[7]{D. Dutta       }
\author[jlab]{D. Gaskell     }
\author[jlab]{O. Hansen      }
\author[9]{F. Hauenstein  }
\author[10]{N. Heinrich   }
\author[jlab]{W. Henry       }
\author[5]{T. Horn        }
\author[10]{G.M. Huber    }
\author[temple]{S. Jia         }
\author[11]{S. Joosten    }
\author[7]{A. Karki       }
\author[10]{S.J.D. Kay    }
\author[10]{V. Kumar      }
\author[16]{X. Li         }
\author[8]{W.B. Li        }
\author[6]{A. H. Liyanage }
\author[jlab]{D. Mack}
\author[jlab]{S. Malace      }
\author[florida]{P. Markowitz   }
\author[jlab]{M. McCaughan   }
\author[11]{Z.-E. Meziani }
\author[12]{H. Mkrtchyan  }
\author[13]{C. Morean     }
\author[5]{M. Muhoza      }
\author[14]{A. Narayan    }
\author[15,18]{B. Pasquini   }
\author[temple]{M. Rehfuss     }
\author[jlab]{B. Sawatzky    }
\author[jlab]{G.R. Smith       }
\author[16]{A. Smith      }
\author[5]{R. Trotta      }
\author[florida]{C. Yero        }
\author[17]{X. Zheng      }
\author[16]{J. Zhou       }


\begin{abstract}

We report new pion electroproduction measurements in the $\Delta(1232)$ resonance, utilizing the SHMS - HMS magnetic spectrometers of Hall C at Jefferson Lab. The data focus on a region that exhibits a strong and rapidly changing interplay of the mesonic cloud and quark-gluon dynamics in the nucleon. The results are in reasonable agreement with models that employ pion cloud effects and chiral effective field theory calculations, but at the same time they suggest that an improvement is required to the theoretical calculations and provide valuable input that will
allow their refinements. The data illustrate the potential of the magnetic spectrometers setup in Hall C towards the study the $\Delta(1232)$ resonance. These first reported results will be followed by a series of measurements in Hall C, that will expand the studies of the $\Delta(1232)$ resonance offering a high precision insight within a wide kinematic range from low to high momentum transfers. 

\end{abstract}


\end{frontmatter}



\section{Introduction}

The first excited state of the nucleon dominates many nuclear phenomena at energies above the pion-production threshold and holds a central role in the physics of the strong interaction. The study of the $N\rightarrow\Delta$ transition form factors (TFFs) has allowed an in-depth exploration of various aspects of the nucleonic structure. Among the early interests in these measurements, one finds the effort to decode the complex quark-gluon and meson cloud dynamics of hadrons that give rise to non-spherical components in their wavefunction, that in a classical limit and at large wavelengths will correspond to a ``deformation"~\cite{Ru75,glashow,soh,revmod}. For the proton, the only stable hadron, the vanishing of the spectroscopic quadrupole moment, due to its spin 1/2 nature, precludes access to the most direct observable of deformation. As a result, the presence of the resonant quadrupole amplitudes $E^{3/2}_{1+}$ and $S^{3/2}_{1+}$ (or E2 and C2 photon absorption multipoles respectively) in the predominantly magnetic dipole $M^{3/2}_{1+}$ (or M1) $\gamma^* N\rightarrow \Delta$ transition emerged as the experimental signature for such an effect \cite{Ru75,glashow,soh, revmod, amb, glas2,capstick,pho2,pho1,pho1b,frol,pos01,merve,bart,Buuren,joo,kun00,Sparveris:2004jn,Kelly:2005jy,kelly, Stave:2006ea,ungaro,Blomberg:2015zma,Blomberg:2019caf,dina,Sato:2000jf,Kamalov:1999hs,Kamalov:2001qg,SAIDweb,Elsner:2005cz,Sparveris:2006uk,longpaper,Aznauryan:2009mx,villano,kirkpatrick,Sparveris:2013ena,quarkpion1,quarkpion2,quarkpion3,pasc,hemmert,hqm,mande:94}. 
The relative strength of the $E2$ and $C2$ amplitudes is normally quoted in terms of their ratio to the dominant magnetic dipole, namely through the $EMR$ and $CMR$ ratio, respectively. 
The TFFs have been explored up to four momentum transfer squared $Q^2=6~(GeV/c)^2$~ \cite{pho2,pho1,pho1b,frol,pos01,merve,bart,Buuren,joo,kun00,joo,Sparveris:2004jn,Kelly:2005jy,kelly, Stave:2006ea,ungaro,Elsner:2005cz,Sparveris:2006uk,longpaper,Aznauryan:2009mx,villano,kirkpatrick,Sparveris:2013ena,Blomberg:2015zma,Blomberg:2019caf}. The results have been found in reasonable agreement with models invoking the presence of non-spherical components in the nucleon wavefunction. Under the prism of the constituent-quark picture of hadrons, these
amplitudes are a consequence of  the non-central, color-hyperfine
interaction among quarks \cite{glashow,glas2}. Nevertheless, this mechanism provides only a fraction of the
observed signal at low momentum transfers. The predicted quadrupole amplitudes~\cite{capstick} are an  order of magnitude smaller compared to the the experimental results and the dominant magnetic dipole amplitude comes $\approx$~30\% short of the experimental measurements. The source for these dynamical
shortcomings can be traced to the fact that such quark models do not respect chiral
symmetry, whose spontaneous breaking leads to strong emission of
virtual pions (Nambu-Goldstone Bosons) \cite{amb}. These couple to
nucleons as $\vec{\sigma}\cdot \vec{p}$ where $\vec{\sigma}$ is the
nucleon spin, and $\vec{p}$ is the pion momentum. The coupling is
strong in the p wave and mixes in non-zero angular momentum
components. Based on this, it is physically reasonable to expect
that the pionic contributions increase the M1 and dominate the E2
and C2 transition matrix elements at low $Q^2$. This has been indicated with the inclusion of pionic effects to quark
models \cite{quarkpion1,quarkpion2,quarkpion3}, in pion
cloud model calculations \cite{Sato:2000jf,Kamalov:1999hs}, and recently
demonstrated in chiral Effective Field Theory ($\chi$EFT) calculations 
\cite{pasc}.

The $\chi$EFT provides a firm theoretical framework at low scales, with
the relevant symmetries of QCD built in consistently. A challenge for the
N to $\Delta$ transition involves the interplay of
two light mass scales, the pion mass and the $N - \Delta$ mass difference.
Studies to consider these two mass scales have been performed
within the framework of heavy-baryon chiral perturbation theory~\cite{Butler:1993ht},
the ``$\epsilon$-expansion'' scheme~\cite{Gellas:1998wx,Gail:2005gz} where
the two pion mass and the $\Delta$-resonance excitation energy scales
are counted as being of the same order, and  
the  ``$\delta$-expansion'' scheme ~\cite{Pascalutsa:2002pi} that provides 
an energy-dependent power-counting scheme that takes into 
account the large variation of the $\Delta$-resonance contributions
with energy, and treats the two light scales $\epsilon$ and $\delta$
on a different footing, counting $\epsilon \sim \delta^2$,
the closest integer-power relation
between these parameters in the real world.

The direct path to calculate the $N$ to $\Delta$
transition form factors starting from the  underlying theory of QCD is provided by Lattice QCD (LQCD).
The LQCD calculations~\cite{dina,lattice-2} have been performed so far with pion mass down to $\sim 300~MeV$, where the $\Delta$ is still stable.
These results tend to somewhat underestimate the M1, similarly to what has been observed in results for the nucleon EM form factors. 
The LQCD results for EMR and CMR ratios on the other hand exhibit remarkable agreement with the experimental measurements, indicating that the ratios are much less affected by lattice artifacts than each of the quantities separately. The statistical uncertainties of the early LQCD results for the two ratios are relatively large due to the fact that the quadrupole amplitudes are sub-dominant and challenging to determine. Progress in recent years enables LQCD calculations to be conducted with physical pion mass, and with statistical uncertainties that are comparable to the experimental ones. Such efforts are currently ongoing, thus making the need for new experimental measurements timely and important. A nice feature of the Lattice QCD calculations is that they have the ability to offer valuable geometrical insight to the nucleon, as illustrated e.g.
through calculations of the three-dimensional contour plot of the $\Delta^+$~\cite{lattice-sh02} and of the $\Delta^+$ quark transverse charge density~\cite{lattice-sh09}.





\begin{figure*}[t]
\centering
\includegraphics[width=18cm]{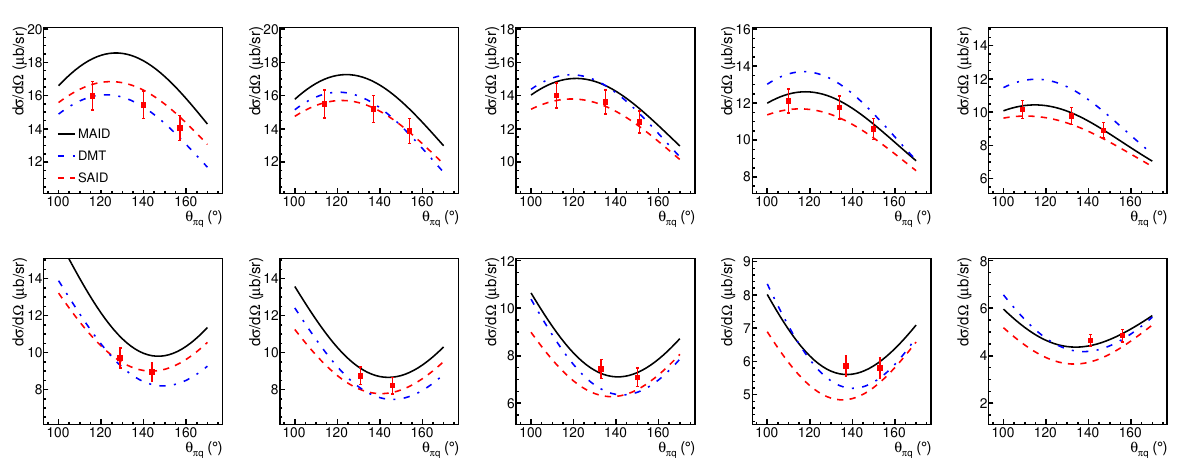}
\caption{Angular dependence of cross section measurements at $Q^2=0.36~GeV^2$ for in-plane kinematics. The top and bottom panels correspond to $\phi_{\pi q}^*=0^\circ$ and $180^\circ$, respectively. From left to right, the results correspond to $W=1212, 1222, 1232, 1242$, and $1252~MeV$, respectively. The data are compared to the theoretical predictions of MAID~\cite{maid}, DMT~\cite{Kamalov:2001qg} and SAID~\cite{SAIDweb}.}
\label{fig-inplane}
\end{figure*}

\section{The experimental measurements}

The reported data were acquired in Hall C of Jefferson Lab during the E12-15-001 experiment. 
For the measurement of the ep$\rightarrow$ep$\pi^\circ$ reaction, electrons with energies of 4.56~GeV at a beam current up to 20~$\mu A$ were produced by Jefferson Lab’s Continuous Electron Beam Accelerator Facility (CEBAF). The electrons were scattered from a 10~cm long liquid-hydrogen target at a temperature of 19 K. The thickness of the aluminum target cell at the entrance and exit is 0.150 (11) mm and 0.191 (19) mm, respectively. 
For every kinematical setting, data were taken with a target made of two aluminum foils located at the positions of the cryotarget entrance 
and exit windows, each having a thickness of 0.6463(10) mm, in order to subtract the background contributions emerging from the target walls by scaling the thicknesses of the two targets.
The scattered electron and recoil proton of the reaction are detected with two magnetic spectrometers, in coincidence. 
The outgoing pion is identified through the reconstructed missing mass spectrum. 
The polar angle $\theta_{\gamma^*\pi}$ of the reaction is defined as the center-of-mass (c.m.) polar angle of the pion with respect to the momentum transfer direction.
The azimuthal angle of the reaction $\phi_{\gamma^*\pi}$ defines the angle between the plane of the two (incoming and scattered) electrons and the pion-proton plane. 
The four-momentum of the outgoing pion, denoted by $\mathbf q'$, is reconstructed as $\mathbf{q'=k+p-k'-p'}$, where $\mathbf{k}$ and $\mathbf{p}$ are the four-momenta of the incoming electron and the target proton, while $\mathbf{k'}$ and $\mathbf{p'}$ are the four-momenta of the final electron and proton, respectively. The four-momentum of the virtual photon is $\mathbf{q=k-k'}$, with $Q^2 \equiv \mathbf{-q^2}$.


The beam properties were monitored throughout the experiment with the Hall C beam diagnostic elements. The beam position monitors (BPMs), that consist of a 4-wire antenna array of open ended thin wire striplines tuned to the RF frequency of the beam, were used to determine the position and the direction of the beam on the experimental target point. The beam current monitors (BCMs), a set of resonant-cavity based beam-current monitors and a parametric current transformer monitor, were used for the continuous non-intercepting beam current measurements. The beam size was measured by using harp scanners, which moved a thin wire through the beam. The beam was rastered over a 2$\times$2 mm$^2$ area to avoid overheating the target. The beam energy was determined with an uncertainty of 0.06\% by measuring the bend angle of the beam, on its way into Hall C, as it traversed the Hall C arc dipole magnets. The total accumulated beam charge was determined with 0.5\% uncertainty. The liquid-hydrogen target density receives contributions from both the target temperature and target boiling effects. The density of the liquid hydrogen target has a nearly linear dependence on the temperature. The temperature is 19 K $\pm$ 0.03 K (intrinsic electronics noise) $\pm$0.05 K (systematic uncertainty), resulting to a target density of 0.0725$\pm$0.0003 $g/cm^3$. For the target boiling effects, a correction was applied to account for the change in the target density caused by beam heating, contributing a density fluctuation of 0.7\% at the maximum current of 20 $\mu A$ used in the experiment. The target length is measured to be 100 $\pm$ 0.26~mm thus resulting to a 0.26$\%$ uncertainty to the cross section measurement.

Two magnetic spectrometers, the Super High Momentum Spectrometer (SHMS) and the High Momentum Spectrometer (HMS) were used to detect, in coincidence, the scattered electrons and recoil protons, respectively. Both spectrometers involve a series of superconducting magnets, including quadrupoles and dipoles, followed by a set of particle detectors. 
The dipole magnets deﬂect charged particles vertically as they enter the detector huts, while the quadrupole magnets optimize the ﬂux of the charged particles entering the dipole magnet and focus the orbits of the charged particles into the detector huts. The two spectrometers are equipped with similar detector packages, with some differentiation due to the different momentum ranges of the spectrometers. The SHMS is also equipped with a Pb-glass calorimeter that can serve as a particle identification detector.  A pair of drift chambers, each with 6 wire planes, separated by about a meter was used to provide the tracking of the detected particles. The uncertainty in the determination of the tracking efficiency was 0.5$\%$ and 1$\%$ for the SHMS and the HMS, respectively.
A set of hodoscope planes was used to form the trigger and to
provide time-of-flight information. The time-of-flight in the HMS spectrometer was used for the proton identification, providing a $>$~20~ns separation from kaons and pions. 
The trigger efficiency of both spectrometer arms is at the 99.9\% level and comes with a $\pm0.1\%$ uncertainty.
For the correction due to the proton absorption in the spectrometer, elastic hydrogen data was taken to determine the fractional loss of protons due to inelastic collisions with material as the proton travelled from the target to the focal plane hodoscope. The fractional loss was determined with an uncertainty of 0.20\%. This correction was applied to the data and the error was included in the systematic uncertainty of the measurement.
The particle tracks are traced, through the spectrometer optics, to the target to provide the particle momentum, scattering angle and target position information. Both spectrometers offer a better than $0.1\%$ momentum resolution and an angular resolution of $\sim$ 1 mrad. The determination of the scattering angle for the SHMS and the HMS spectrometers comes with a 0.5~mrad uncertainty that is determined from constraints on the elastic kinematic reconstruction.

\begin{figure*}[t]
\centering
\includegraphics[width=18cm]{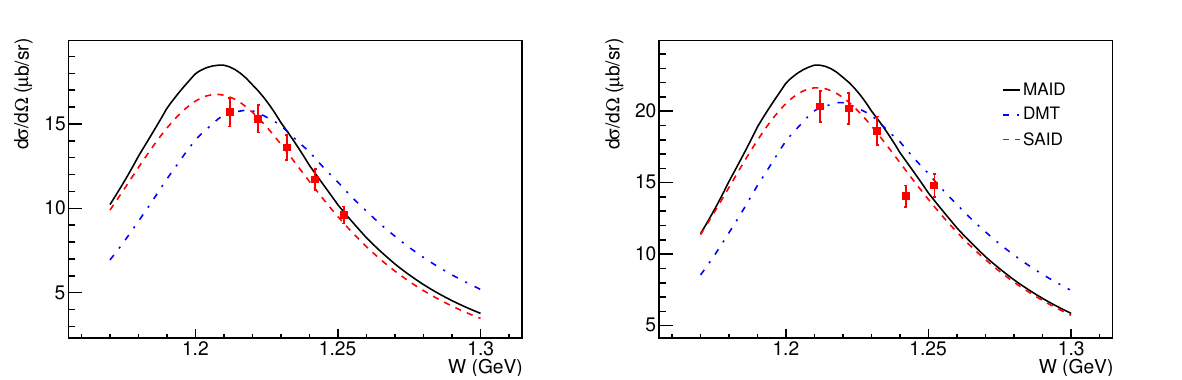}
\caption{The W-dependence of the cross section measurements at $Q^2=0.36~GeV^2$. The left panel corresponds to $\phi_{\pi q}^*=0^\circ$ and $\theta_{\pi q}^{*}=135^\circ$, and the right panel to $\phi_{\pi q}^*=35^\circ$ and $\theta_{\pi q}^{*}=130^\circ$ . The data are compared to the theoretical predictions of MAID~\cite{maid}, DMT~\cite{Kamalov:2001qg} and SAID~\cite{SAIDweb}.}
\label{fig-w}
\end{figure*}

The coincidence time was determined as the difference in the time-of-flight between the two spectrometers, accounting for path-length variation corrections from the central trajectory and for the individual start-times. The experimental setup provided a better than 1~ns (FWHM) resolution in the coincidence timing spectrum that was measured within an 80~ns timing window. Random coincidences were subtracted using the accidental bands of the coincidence time spectrum. The uncertainty to the live-time correction, that accounts for the electronics and computer dead-time, ranged between 0.3$\%$ and 0.6$\%$ for the different kinematic settings of the experiment. To estimate the systematic error on this correction, we used the standard deviation of the Gaussian fit to the histogram of the deadtime of the runs used in each kinematic setting. The duration of each run was typically about half an hour of beam time, and the number of runs per kinematic setting ranged from about 50 to 100.

The events of the exclusive reaction ep$\rightarrow$ep$\pi^\circ$ were identified from the missing-mass reconstruction, through a selection cut around the photon peak in the missing-mass-squared spectrum. The true momentum settings of the two spectrometers were determined based on a cross-calibration method that utilizes pairs of azimuthal asymmetry measurements. Here, the momentum and position of the electron spectrometer remain the same between the two kinematical settings. The momentum setting for the proton spectrometer also remains constant, while the proton spectrometer is re-positioned symmetrically with respect to the momentum transfer direction. Since the two kinematical settings involve identical momentum settings for each of the two spectrometers, the determination of their absolute momentum settings comes from a unique solution for both kinematics, that simultaneously calibrates the reconstructed missing mass peak to the physical value of the pion mass. Following the above procedure, the correction between the set and the true values in the central momentum of the two spectrometers was determined to be smaller than 0.1$\%$.

To determine the stability over time as well as the proper normalization, elastic scattering measurements with a proton target were performed throughout the experiment. The results are stable and consistent, within the experimental uncertainties, with the world elastic data. This points out to a consistency in the control of luminosity, target density and beam position, along with the ability to position the spectrometers reliably in the experimental hall and to consistently set and control their central momenta.



\section{Results and discussion}

The five-fold differential cross section for the
$p(e,e'p)\pi^{0}$ reaction is written as a sum of two-fold
differential cross sections with an explicit $\phi^*$
dependence as
\begin{eqnarray}
\displaystyle \frac{d^5\sigma}{d\Omega_e d\Omega^*_{\pi} d\omega}  = & \Gamma (\sigma_T + \epsilon
\sigma_L + v_{LT}\sigma_{LT}\cos\phi_{\pi q}^*  \nonumber \\
  &  \displaystyle + ~ \epsilon \sigma_{TT} \cos 2\phi_{\pi q}^*)
\label{eq:XS}
\end{eqnarray}
where $\phi_{\pi q}^*$ is the pion center of mass azimuthal angle with
respect to the electron scattering plane, $v_{LT}=\sqrt{2\epsilon(1+\epsilon)}$, $\epsilon$ is the transverse polarization of the virtual photon,
and $\Gamma$ is the virtual photon flux.
The differential cross sections
($\sigma_{T},\sigma_{L},\sigma_{LT},\sigma_{TT}$) 
are all functions of the
center of mass energy $W$, the four momentum transfer squared $Q^2$,
and the pion center of mass polar angle $\theta_{\pi q}^{*}$ and they are bilinear combinations
of the multipoles. The E2 and C2 amplitudes manifest themselves mostly
through the interference with the dominant dipole (M1) amplitude.
The longitudinal-transverse (LT) response is sensitive to
the C2 amplitude through the interference of the C2 amplitude
with the M1, while the transverse-transverse (TT) response is
sensitive to the E2 amplitude through the interference of the
E2 amplitude with the M1. The $\sigma_T + \epsilon
\sigma_L$ partial cross section is
dominated by the M1 multipole.

\begin{figure*}[t]
\centering
\includegraphics[width=18cm]{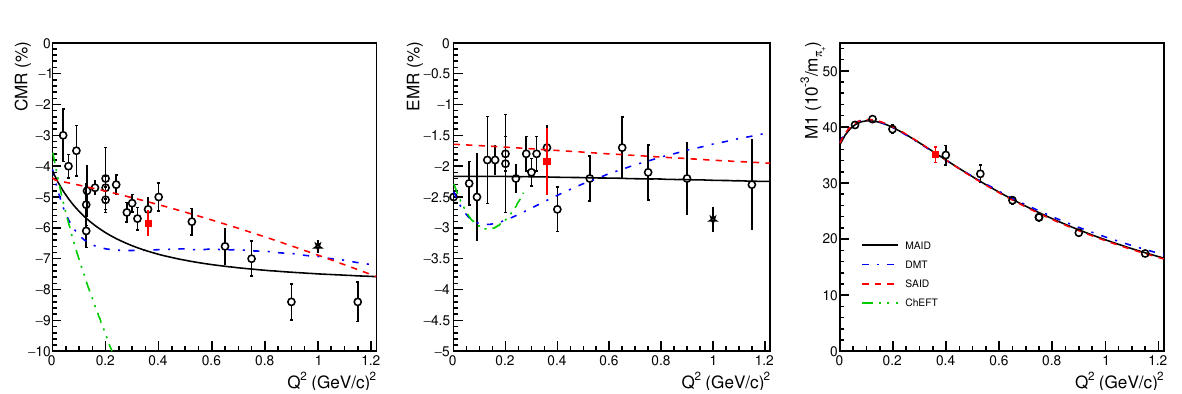}
\caption{The extracted quadrupole and dipole transition form factors (filled boxes). The world data~ \cite{pho2,pho1,pho1b,frol,pos01,merve,bart,Buuren,joo,kun00,joo,Sparveris:2004jn,Kelly:2005jy,kelly, Stave:2006ea,Blomberg:2015zma,Blomberg:2019caf,Elsner:2005cz,Sparveris:2006uk,longpaper,Aznauryan:2009mx,Sparveris:2013ena} are shown with open symbols. The Hall A measurement at $Q^2=1~GeV^2$~\cite{Kelly:2005jy,kelly} is shown as a star, to distinguish from the CLAS measurements (open-circles) in the surrounding region. The data are compared to the theoretical predictions of MAID~\cite{maid}, DMT~\cite{Kamalov:2001qg}, SAID~\cite{SAIDweb} and the ChEFT calculation~\cite{pasc}.}
\label{fig-multi}
\end{figure*}

For the measurement of the cross section, the determination of the coincidence acceptance is calculated with the Hall
C Monte Carlo simulation program, SIMC, which integrates the beam configuration, target geometry, spectrometer acceptances, resolution effects, energy losses and radiative corrections. The cross section is first averaged over the multidimensional phase space within the measured analysis bin, and is then followed by a kinematic translation procedure, namely bin centering corrections, that converts the cross section that has been averaged over finite phase space to a final point cross section extracted at the central kinematic
values of the phase space. For that part, theoretical predictions from various models are integrated in the simulation of the experiment and are studied over the same volume in phase space as the data. The bin centering corrections are small, typically 2\% to
3\%, indicating that the cross section tends to vary smoothly
and fairly symmetrically through the phase space. The systematic uncertainty to this correction is studied by employing different theoretical models as well as by applying variations to the size of the analysis bins, and has been found to be small compared to the experimental uncertainties.

The measurements were conducted at intermediate momentum transfer kinematics of $Q^2=0.36~GeV^2$. Cross sections were measured within a $W$ range from 1210~$MeV$ to 1250~$MeV$, with an extended coverage in the polar angle $\theta_{\pi q}^{*}$, and a reach in the azimuthal angle $\phi_{\pi q}^*$ that extends from in-plane kinematics up to 50$^\circ$ out-of-plane angles. A subset of the measured cross sections, for the in-plane kinematics, is shown in Fig.~\ref{fig-inplane}. The data are compared to the theoretical predictions of MAID~\cite{maid}, DMT~\cite{Kamalov:2001qg} and SAID~\cite{SAIDweb}. The MAID and SAID calculations are primarily phenomenological, while the DMT contains explicit pion cloud contributions. An observation is that while the models follow a similar $\theta_{\pi q}^{*}$ dependence, they tend to disagree with  each other in absolute magnitude, and occasionally with the data across the resonance region. Fig.~\ref{fig-w} gives an insight to the $W$ dependence of the measured cross section. The MAID prediction tends to overestimate the measured cross sections at the lower wing of the resonance, similarly to what has been observed in previous measurements lower than $Q^2=0.2~GeV^2$~\cite{longpaper,Sparveris:2013ena}.
Overall, improvements are in order for all the models, and the reported measurements provide new input and guidance towards this direction. The reported measurements are summarized in Table~\ref{tab:cross1} and Table~\ref{tab:cross2}. Fits of the resonant amplitudes have been performed at $Q^2=0.36~GeV^2$ while taking into account the background amplitude contributions from MAID and DMT. In these fits, the differences between the model descriptions of the background terms
results in a deviation of the fitted amplitudes, which is indicative of the level of the model uncertainty. We find that $CMR=(-5.85 \pm 0.28_{exp} \pm 0.20_{mod})\%$ and $EMR=(-1.93\pm 0.50_{exp} \pm 0.10_{mod})\%$.

The extracted quadrupole and magnetic dipole amplitudes are in good agreement with the trend of the world data and they deviate considerably from the Constituent quark model (CQM) predictions e.g.~\cite{capstick,hqm}, reconfirming that the color hyperfine interaction is inadequate to explain the effect at large distances. A more meaningful comparison is provided by the theoretical model predictions from MAID~\cite{maid}, DMT~\cite{Kamalov:2001qg}, SAID~\cite{SAIDweb}, and the ChEFT calculation~\cite{pasc}, as shown in 
Fig.~\ref{fig-multi}. For the ChEFT~\cite{pasc}, an estimate of the model uncertainty is derived by calculating the magnitude of the next order terms in
the chiral expansion. This results to a theoretical uncertainty of $\sim \pm 1\%$ and $\pm 2 \%$ for the EMR and the CMR ratios, respectively in the region around $Q^2=0.2~GeV^2$. The calculation is solidly based on QCD and successfully accounts for the magnitude of the effects for the EMR, while for the CMR a rapid divergence from the experimental measurements is observed above $Q^2=0.2~GeV^2$. The ChEFT calculation gives overall credence to the dominance of the meson cloud, nevertheless, the size of the theoretical uncertainties make the need for the next order calculation obvious. The reported data overlap with the low-$Q^2$ domain of the CLAS measurements~\cite{Aznauryan:2009mx} and confirm their findings. The data illustrate the potential of employing the experimental setup in Hall C for the study of the $\Delta(1232)$ resonance. A series of follow up experiments using the same experimental setup has been approved at JLab, and will expand these studies with high precision measurements within a wide kinematic range from $Q^2=0.01~GeV^2$ to $0.7~GeV^2$. At low momentum transfers, they will allow an in-depth study of the mesonic cloud dynamics in a region were they are dominant and will provide a stringent test to the QCD prediction that the two quadrupole amplitudes converge at $Q^2\rightarrow 0$~\cite{proposal-lowq}. At higher $Q^2$, the CLAS data suggest a steeper fall-off for the CMR compared to the findings of the high precision recoil polarization measurement of Hall A at  $Q^2=1~GeV^2$~\cite{Kelly:2005jy,kelly}, as seen in Fig.~\ref{fig-multi}. Here, the upcoming measurements in Hall C will come to complement the CLAS data, adding to our understanding of the high-$Q^2$ dependence of the transition form factors.

In conclusion, we present cross section measurements of the $\pi^\circ$ electroproduction reaction in the $\Delta(1232)$ resonance region, at intermediate momentum transfer kinematics of $Q^2=0.36~GeV^2$. The data provide a precise determination of the two quadrupole and of the magnetic dipole $N\rightarrow\Delta$ transition form factors. The cross section measurements are found in reasonable agreement with theoretical calculations that include pion cloud contributions and with ChEFT calculations. At the same time, they indicate that some improvement is required to the theoretical calculations and they provide valuable input that will allow their refinements, thus offering valuable input towards the understanding of the nucleon dynamics. 

We would like to thank the JLab Hall C technical staff and the Accelerator Division for their outstanding support. This
work has been supported by the US Department of Energy Office of Science, office of Nuclear Physics under contract
no. DE-SC0016577.

\section*{References}

\bibliographystyle{elsarticle-num}

\onecolumn


\begin{table}[htbp]
    \centering
     \caption{Cross section measurements at $W=1212, 1222$ and $1232~MeV$.}
    \resizebox{0.55\textwidth}{!}{
    \renewcommand{\arraystretch}{1.5}
    \begin{tabular}{|c|c|c|c|}
        \hline
        \textbf{W ($GeV$)} & \textbf{$\phi^*_{\pi q}$ ($deg$)} & \textbf{$\theta^*_{\pi q}$ ($deg$)} & \textbf{$\sigma$ $\pm$ $\delta \sigma_{stat}$ $\pm$ $\delta \sigma_{sys}$ ($\mu b/sr$)} \\
        \hline
        \multirow{5}{*}{1.212} &0 & 116 & $15.99 \pm 0.11 \pm 0.84$ \\
         &0 & 135 & $15.69 \pm 0.12 \pm 0.83 $ \\
         &0 & 140 & $15.44 \pm 0.12 \pm 0.82$ \\
         &0 & 157 & $14.05 \pm 0.13 \pm 0.74$ \\
         & 20 & 115 & $18.01 \pm 0.14 \pm 0.95$ \\
         & 30 & 136 & $18.51 \pm 0.16 \pm 0.98$ \\
         & 35 & 130 & $20.33 \pm 0.18 \pm 1.08$ \\
         & 40 & 150 & $16.75 \pm 0.16 \pm 0.89$ \\
         & 160 & 130 & $10.48 \pm 0.25 \pm 0.55$ \\
         & 160 & 144 & $9.52 \pm 0.11 \pm 0.50$ \\
         &180 & 144 & $8.95 \pm 0.09 \pm 0.47$ \\
         &180  & 129 & $9.70 \pm 0.16 \pm 0.51$ \\
        \hline
        \multirow{5}{*}{1.222} &0 & 114 & $15.49 \pm 0.09 \pm 0.82$ \\
         &0 & 135 & $15.30 \pm 0.11 \pm 0.81$ \\
         & 0 & 154 & $13.86 \pm 0.12 \pm 0.73$ \\
         & 20 & 112 & $17.92 \pm 0.13 \pm 0.95$ \\
         & 30 & 134 & $18.42 \pm 0.15 \pm 0.97$ \\
         & 35 & 130 & $20.20 \pm 0.16 \pm 1.06$ \\
         & 45 & 146 & $18.32 \pm 0.16 \pm 0.97$ \\
         & 155 & 144 & $9.91 \pm 0.08 \pm 0.52$ \\
         & 160 & 130 & $10.26 \pm 0.12 \pm 0.54$ \\
         &180  & 131 & $8.73 \pm 0.08 \pm 0.46$ \\
         &180 & 146 & $8.21 \pm 0.06 \pm 0.43$ \\
    
        \hline
        \multirow{5}{*}{1.232} &0 & 112 & $14.00 \pm 0.08 \pm 0.74$ \\
         &0 & 135 & $13.61 \pm 0.09 \pm 0.72$ \\
         &0 & 151 & $12.41 \pm 0.10 \pm 0.66$ \\
         & 20 & 112 & $16.67 \pm 0.12 \pm 0.88$ \\
         & 35 & 130 & $18.64 \pm 0.15 \pm 0.98$ \\
         & 50 & 142 & $18.68 \pm 0.16 \pm 0.99$ \\
         & 140 & 141 & $11.88 \pm 0.10 \pm 0.63$ \\
         & 155 & 130 & $10.20 \pm 0.10 \pm 0.54$ \\
         &180  & 133 & $7.45 \pm 0.06 \pm 0.39$ \\
         &180 & 150 & $7.09 \pm 0.06 \pm 0.38$ \\

        \hline
    \end{tabular}%
     }
    \label{tab:cross1}%
    
\end{table}%


\begin{table}[htbp]
    \centering
     \caption{Cross section measurements at $W=1242$ and $1252~MeV$.}
    \resizebox{0.55\textwidth}{!}{
    \renewcommand{\arraystretch}{1.5}
    \begin{tabular}{|c|c|c|c|}
        \hline
        \textbf{W ($GeV$)} & \textbf{$\phi^*_{\pi q}$ ($deg$)} & \textbf{$\theta^*_{\pi q}$ ($deg$)} & \textbf{$\sigma$ $\pm$ $\delta \sigma_{stat}$ $\pm$ $\delta \sigma_{sys}$ ($\mu b/sr$)} \\
        \hline
     
        \multirow{5}{*}{1.242} &0 & 110 & $12.11 \pm 0.07 \pm 0.64$ \\
         &0 & 135 & $11.71 \pm 0.08 \pm 0.62$ \\
         &0 & 150 & $10.61 \pm 0.08 \pm 0.56$ \\
         & 20 & 108 & $17.06 \pm 0.13 \pm 0.90$ \\
         & 35 & 130 & $14.07 \pm 0.10 \pm 0.74$ \\
         & 50 & 138 & $17.40 \pm 0.14 \pm 0.92$ \\
         & 140 & 143 & $10.01 \pm 0.09 \pm 0.53$ \\
         & 155 & 133 & $8.66 \pm 0.08 \pm 0.46$ \\
         &180  & 137 & $5.87 \pm 0.05 \pm 0.31$ \\
         &180 & 153 & $5.80 \pm 0.06 \pm 0.31$ \\
        
        \hline
        \multirow{5}{*}{1.252} &0 & 109 & $10.16 \pm 0.06 \pm 0.54$ \\
         &0 & 132 & $9.77 \pm 0.06 \pm 0.52$ \\
         &0 & 135 & $9.60 \pm 0.06 \pm 0.51$ \\
         &0& 147 & $8.91 \pm 0.07 \pm 0.47$ \\
         & 20 & 107 & $12.40 \pm 0.08 \pm 0.66$ \\
         & 35 & 125 & $15.70 \pm 0.13 \pm 0.83$ \\
         & 35 & 130 & $14.81 \pm 0.12 \pm 0.78$ \\
         & 50 & 134 & $7.91 \pm 0.08 \pm 0.42$ \\
         & 135 & 144 & $14.74 \pm 0.11 \pm 0.78$ \\
         & 150 & 135 & $9.03 \pm 0.08 \pm 0.48$ \\
         &180  & 141 & $4.65 \pm 0.05 \pm 0.25$ \\
         &180 & 156 & $4.84 \pm 0.05 \pm 0.26$ \\
         
        \hline
    \end{tabular}%
     }
    \label{tab:cross2}%
   
\end{table}%



\end{document}